\documentclass[apjl,tighten]{emulateapj}
\usepackage{apjfonts}
\usepackage{longtable}
\usepackage{ulem}
\usepackage{verbatim}
\newcommand{\zsun}{$Z_\odot$}
\newcommand{\msun}{$M_\odot$}

\newcommand{\hi}{H\,{\sc i}\rm}
\newcommand{\hii}{H\,{\sc ii}\rm}

\newcommand{\nii}{[N\,{\sc ii}]}

\newcommand{\oiii}{[O\,{\sc iii}]}

\newcommand{\sii}{[S\,{\sc ii}]}

\newcommand{\eg}{e.g.~}

\newcommand{\hbeta}{H$\beta$}
\newcommand{\halpha}{H$\alpha$}
\newcommand{\lin}{$\,\lambda$}
\newcommand{\llin}{$\,\lambda\lambda$}

\newcommand{\rtf}{$R_{25}$}

\newcommand{\oh}{12\,+\,log(O/H)}

\newcommand{\vs}{vs.}
\newcommand{\ngc}{NGC~404}

\slugcomment{}

\shorttitle{The rejuvenation of the S0 galaxy NGC~404}
\shortauthors{Bresolin}

\begin{document}

\title{Clues on the rejuvenation of the S0 galaxy NGC~404 from the chemical abundance of its outer disk} 

\author{Fabio Bresolin} \affil{Institute for Astronomy, 2680 Woodlawn
Drive, Honolulu, HI 96822 \\bresolin@ifa.hawaii.edu}

\begin{abstract}
The oxygen abundance of the outer disk of the nearby S0 galaxy \ngc, a  prototypical
early-type galaxy with extended star formation, has been derived
from the analysis of \hii\ region spectra. The  high mean value found, 
\oh\,=\,$8.6 \pm 0.1$, equivalent to approximately 80\% of the solar value, argues against both the previously proposed cold accretion and recent merger scenarios as viable mechanisms for the assembly of the star-forming gas. 
The combination of the  present-day gas metallicity with the published star formation history of this galaxy favors a model in which the recent star forming activity represents the declining tail of the original one.

\end{abstract}

 
%==========================================================================
\section{Introduction}

Studies of the gaseous content of  early-type (E/S0) galaxies (ETGs) have  established that gas accretion and the presence of \hi\ reservoirs are common features among these systems (\citealt{Oosterloo:2010, Davis:2011, Thom:2012}). The presence of gas
is often accompanied by low-level, recent (age $<1$~Gyr) star formation, which is detected in a significant fraction ($> 30\%$) of the ETGs observed in the far-UV by the Galaxy Evolution Explorer (GALEX, \citealt{Martin:2005})  (\citealt{Yi:2005, Kaviraj:2007}).
Several authors (\eg \citealt{Serra:2008,Kaviraj:2009,Cortese:2009})
have invoked the supply of external gas via galaxy merging  or  interactions
in order to explain the star forming and gas content properties of ETGs, as well as the  rejuvenation (\citealt{Rampazzo:2007}) of these systems, that had the bulk of their star formation virtually completed early  in their history.

Star formation  in ETGs is often confined to their outskirts, in the form of extended UV-bright ring-like structures (\citealt{Donovan:2009, Thilker:2010, Salim:2010, Moffett:2012}), resembling the outer star-forming  disks identified in late-type galaxies (\citealt{Thilker:2007}). Despite their prominence in far-UV GALEX images of nearby ETGs, these rings account for only a few percent of the total stellar mass of the host galaxies (\citealt{Marino:2011}).

In a recent study of 29 S0 galaxies at redshift $z\sim 0.1$ 
\citet{Salim:2012} critically reviewed the  mechanisms that can generate spatially extended star formation in ETGs. Cold, smooth gas accretion from the intergalactic medium (IGM), as predicted by cosmological hydrodynamic simulations (\eg \citealt{Keres:2009b}), was identified, based on qualitative optical/UV galaxy morphology,
 as the likely mechanism for rejuvenating the star formation activity in about half of their sample. The supply of gas from minor mergers and the fading of the original star formation (not requiring external gas sources)
could explain the remaining cases. These conclusions have been strengthened in a related work by \citet{Fang:2012}, who included 670 extended star formation ETG candidates in their quantitative analysis of UV-optical colors. Thus, ETGs with extended star formation probably constitute   a heterogenous class, composed both by systems with declining star formation and others experiencing  rejuvenation following gas accretion.

The nearest S0 galaxy, NGC~404 (D\,=\,3.05~Mpc, \citealt{Dalcanton:2009}) has been identified by \citet{Thilker:2010} as the host of recent star formation in its nearly face-on outer disk. UV-emitting young star clusters reside within an \hi\ ring, extending between 1.5 and 6.5 kpc (1\,--\,4\,\rtf) from the galactic center.
The relatively small distance to NGC~404 and its isolation (\citealt{Karachentsev:2002a}) make this galaxy an ideal site where to explore the processes  regulating the evolution of an ETG with extended star formation. 
Following the suggestion by \citet{del-Rio:2004} that a merger with a dwarf galaxy supplied NGC~404 with fresh gas in the past Gyr, \citet{Thilker:2010} 
argued that this lenticular galaxy is an example of a rejuvenated ETG, currently shifting from the red sequence of the bimodal  optical color distribution of galaxies back into the green valley.
This paper offers an alternative view, providing constraints on the nature of the extended star formation in NGC~404 from a chemical abundance analysis of \hii\ regions lying in its outer disk.

% - - - - - - - - - - - - - - - - - - - - - - - - - - - - - - - - - - - - - - - - - - - - - -
\section{Observations and data reduction}

Observations of \hii\ region candidates in the outer disk of \ngc\ were obtained with the Gemini Multi-Object Spectrograph (GMOS, \citealt{Hook:2004}) at the Gemini North telescope. The targets were selected from \halpha\ narrow-band images of two 5\farcm5$\times$5\farcm5 GMOS fields in the outer, star-forming disk of the  galaxy,  approximately 3\farcm9 (3.4~kpc) E and 4\farcm6 (4.1~kpc)  N of the center, respectively.
Spectra were acquired  on September 18, 2012, under $0\farcs5$ FWHM seeing conditions, using two multi-object masks, one per field, containing 1\farcs5-wide slits. Three 1800\,s exposures were secured for each field using the B600 grating,  providing spectra covering 
the $\sim$4700-7500\,\AA\ wavelength range (the  coverage depending on the spatial distribution of the targets), at a spectral resolution of 6\,\AA\ for nebulae filling the slit (3.5\,\AA\ for six spatially unresolved objects).

\begin{figure}
\medskip
\center \includegraphics[width=1.01\columnwidth]{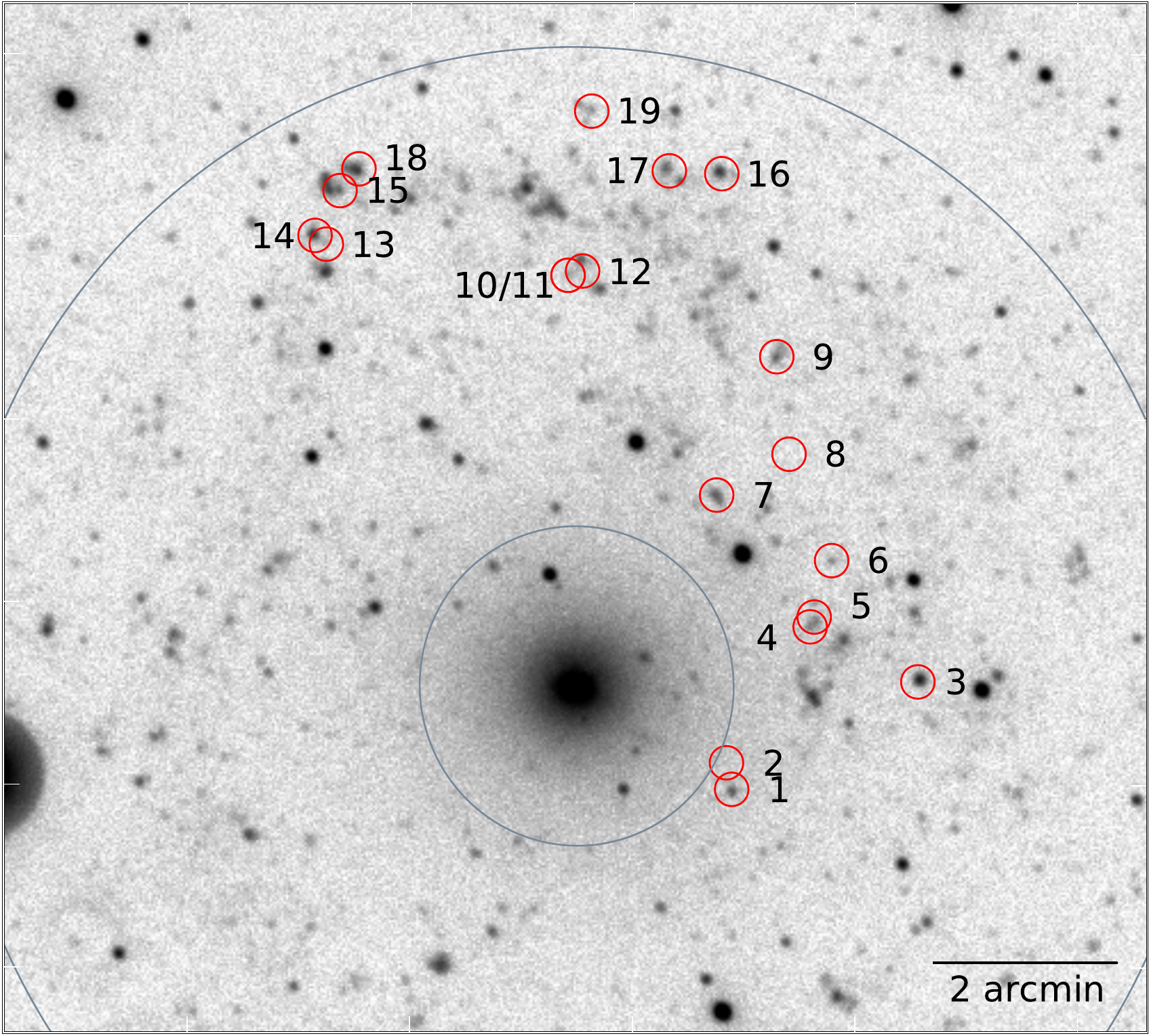}\medskip
\caption{Identification of the  targets on a near-UV GALEX  image of \ngc\ (N at top, E on the left). The circles, drawn at radii \rtf\ and 4\,\rtf, represent the approximate inner and outer boundaries of the star forming ring. \label{image}}
\end{figure}

\begin{deluxetable*}{ccccccccc}

\tablecolumns{9}
\tablewidth{0pt}
\tablecaption{\hii\ region sample: coordinates and line ratios\label{table}}

\tablehead{
\colhead{ID}	     &
\colhead{R.A.}	 &
\colhead{DEC}	 &
\colhead{$R$}		&
\colhead{$R$/\rtf}	 &
\colhead{\oiii\lin 5007/\hbeta}  &
\colhead{\nii\lin 6583/\halpha}	 &
\colhead{\sii\lin 6717,31/\halpha}	&
\colhead{Comments}\\[0.5mm]
\colhead{}       &
\colhead{(J2000.0)}       &
\colhead{(J2000.0)}       &
\colhead{(arcsec)}       &
\colhead{}   &
\colhead{}   &
\colhead{}   &
\colhead{}   &
\colhead{}  }
\startdata
\\[-2mm]
 1 & 01 09 18.65  & 35 41 57.5  &  124 &  1.18 &  $<$ 0.30  &    0.24  &    0.31  &       \\  %b_11
 2 & 01 09 18.94  & 35 42 14.9  &  111 &  1.06 &   7.87  &    0.04  &   $<$\,0.02  &     PN candidate  \\  %b_09
 3 & 01 09 08.62  & 35 43 08.2  &  224 &  2.14 &    \nodata        &    0.20  &    0.18  &       \\  %b_08
 4 & 01 09 14.43  & 35 43 44.3  &  158 &  1.51 &  $<$ 0.36  &    0.23  &    0.28  &       \\  %b_06
 5 & 01 09 14.21  & 35 43 50.7  &  162 &  1.55 &   4.87  &    0.42  &    0.91  &     SNR candidate  \\  %b_07
 6 & 01 09 13.27  & 35 44 27.8  &  187 &  1.78 &   0.94  &    0.22  &    0.24  &       \\  %b_05
 7 & 01 09 19.48  & 35 45 10.8  &  157 &  1.49 &   1.20  &    0.21  &    0.17  &       \\  %b_04
 8 & 01 09 15.58  & 35 45 37.7  &  208 &  1.98 &  $<$ 0.48  &    0.15  &   $<$\,0.07  &     unresolved  \\  %b_03
 9 & 01 09 16.24  & 35 46 41.7  &  256 &  2.44 &   0.15  &    0.23  &    0.19  &       \\  %b_01
10 & 01 09 27.52  & 35 47 35.0  &  275 &  2.62 &    \nodata        &    0.29  &    0.13  &     unresolved  \\  %a_03_1
11 & 01 09 27.52  & 35 47 35.0  &  275 &  2.62 &    \nodata        &    0.20  &    0.33  &     unresolved  \\  %a_03_2
12 & 01 09 26.73  & 35 47 37.9  &  278 &  2.64 &    \nodata        &    0.24  &    0.28  &       \\  %a_04
13 & 01 09 40.57  & 35 47 55.2  &  339 &  3.23 &    \nodata        &    0.21  &    0.14  &     unresolved  \\  %a_08
14 & 01 09 41.18  & 35 48 01.0  &  348 &  3.31 &   3.10  &    0.10  &    0.09  &       \\  %a_10
15 & 01 09 39.84  & 35 48 30.5  &  367 &  3.50 &   0.15  &    0.17  &    0.22  &       \\  %a_07
16 & 01 09 19.22  & 35 48 41.9  &  355 &  3.38 &   0.56  &    0.17  &    0.22  &       \\  %a_01
17 & 01 09 22.06  & 35 48 43.8  &  350 &  3.33 &  $<$ 0.28  &    0.21  &    0.26  &       \\  %a_02
18 & 01 09 38.82  & 35 48 44.8  &  375 &  3.57 &   0.52  &    0.22  &    0.17  &     unresolved  \\  %a_09
19 & 01 09 26.25  & 35 49 22.9  &  385 &  3.66 &  $<$ 0.43  &    0.21  &    0.27  &      %a_05
\enddata
\end{deluxetable*}

\begin{figure*}
\center \includegraphics[width=2.05\columnwidth]{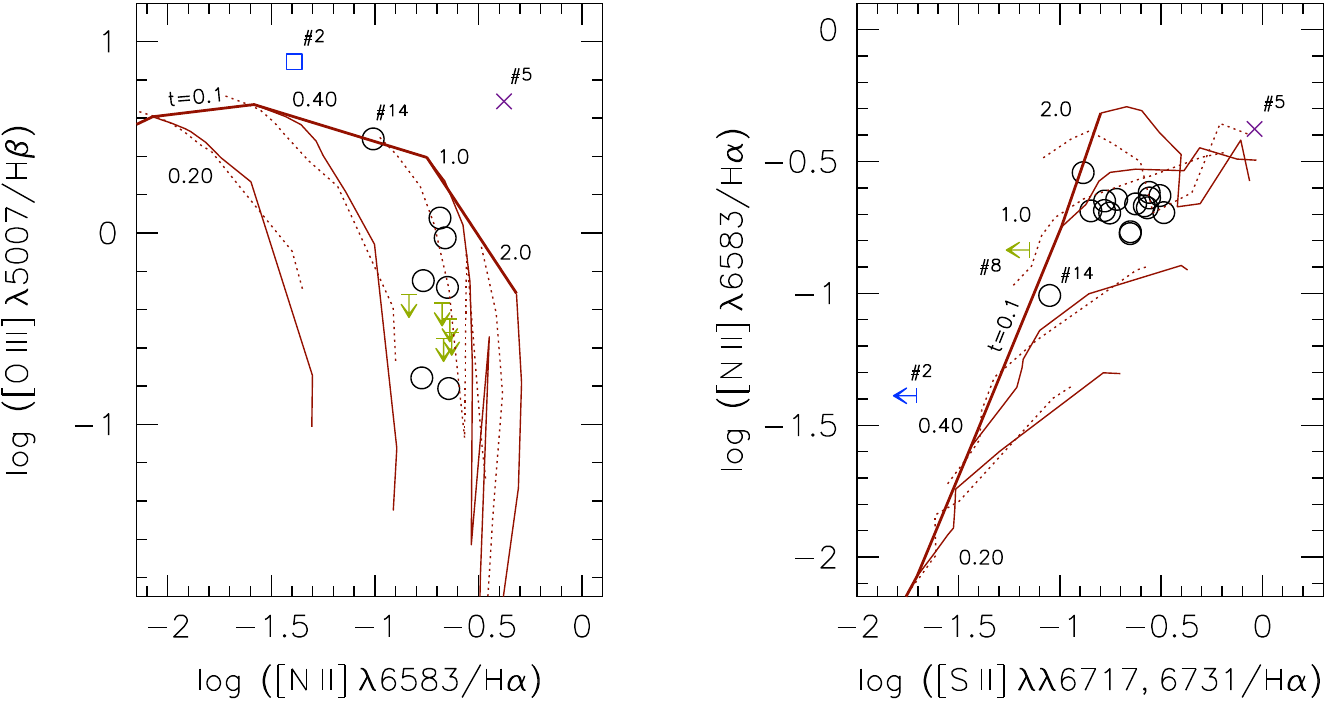}\medskip
\caption{Excitation diagrams: \nii/\halpha\ \vs\ \oiii/\hbeta\, {\it (left)} and \sii/\halpha\ \vs\ \nii/\halpha\ {\it (right)}. Open circles: measured flux ratios. The candidate planetary nebulae  (\#2, \#8) and  supernova remnant (\#5) are identified. The arrows represent upper limits. The curves show theoretical models from \citet{Dopita:2006a} calculated for ages  $0.1-4$~Myr, metallicities 0.2, 0.4, 1.0, 2.0 $\times$~\zsun, $\log\mathcal{R}=-2$ (solid lines) and $\log\mathcal{R}=0$ (dotted lines).
\label{f2}}
\bigskip

\end{figure*}

{\sc iraf}\footnote{{\sc iraf} is distributed by the National Optical Astronomy Observatories, which are operated by the Association of Universities for Research in Astronomy, Inc., under cooperative agreement with the National Science Foundation.}
routines in the {\tt\small gemini/gmos} package were used  for electronic bias subtraction, flat field correction, wavelength calibration, and coaddition of the raw data frames. The spectra were not flux calibrated. This has virtually no effect on the analysis presented here, given the proximity in wavelength between the emission lines used to construct the  nebular diagnostics.
Two of the targets were found to be background emission-line galaxies, and were removed from the  analysis.
The final sample comprises 19 objects in an annulus between \rtf\ and 4\,\rtf\ (hereafter referred to as the star-forming ring) of \ngc. Their location is indicated in Fig.~\ref{image} (\#10 and \#11 are targets included in the same slit). Celestial coordinates are summarized in Table~\ref{table}, together with  galactocentric distances (in arcsec and in units of the isophotal radius \rtf\,=\,$1\farcm75$), 
deprojected adopting the geometrical parameters from \citet{del-Rio:2004}.
\bigskip
\bigskip

\section{Chemical abundance analysis}
The  emission lines covered in the observed spectral range include \halpha, \nii\lin6583 and
\sii\llin6717,\,6731 (the latter remained undetected in \#2 and \#8). For 14 targets the \hbeta\ and \oiii\lin5007 lines were also accessible. 
Table~\ref{table} reports the values (or upper limits) of the line ratios \oiii\lin5007/\hbeta, \nii\lin6583/\halpha\ and \sii\llin6717,\,6731/\halpha, which are virtually independent of flux calibration and reddening corrections, given the small separation in wavelength of the  lines involved.

Fig.~\ref{f2} illustrates the excitation properties of the targets, showing their location in the \nii/\halpha\ \vs\ \oiii/\hbeta\ and \sii/\halpha\ \vs\ \nii/\halpha\   diagrams.
Three objects stand out as peculiar based on their line ratios: \#2 and \#8, classified as candidate 
planetary nebulae (consistent with the point-source appearance and the lack of a continuum), and \#5 
(a 58\,pc-diameter ring identified as a supernova remnant). The rest of the sample  is composed of bona fide \hii\ regions, whose chemical analysis can be carried out using standard  diagnostics. 

The diagrams in Fig.~\ref{f2} include curves representing \hii\ region models from \citet{Dopita:2006a}, calculated for ionizing cluster ages between 0.1 and 4~Myr, and different metallicities (0.2, 0.4, 1.0 and 2.0 $\times$~\zsun).
The parameter $\mathcal{R} \propto M_{cl}/P_0$ (the ratio between cluster mass and pressure of the interstellar medium) 
is also varied between $\log\mathcal{R}=-2$ (solid lines) and $\log\mathcal{R}=0$ (dashed lines), although this has only a secondary effect on the interpretation of the diagrams. According to these models, the bulk of the \hii\ region sample has a metallicity slightly below solar. Since the model curves in the \nii/\halpha\ \vs\ \oiii/\hbeta\, diagram become virtually vertical after a cluster age of 3~Myr, the same conclusion can also be drawn  for the 5 objects having only an upper limit for the \oiii/\hbeta\ ratio.

In order to quantify the radial distribution of the nebular metallicities the abundance diagnostic N2 = log(\nii\lin6583/\halpha)
was adopted, using the  calibrations provided by \citet[=\,PP04]{Pettini:2004} and \citet[=\,D02]{Denicolo:2002}.
The resulting O/H abundances are presented in Fig.~\ref{f3}, using solid  (PP04) and open (D02) circles. While most sources share a similar O/H ratio, \#14 (identified in the figure) displays a significantly lower abundance. This object has the largest \halpha\ equivalent width (950\,\AA) in the sample, indicating a very young nebula, which is supported by the models in Fig.~\ref{f2} (point lying on the 0.1~Myr curve). The 
position of \#14 in relation to the model curves also shows that, despite the low \nii/\halpha\ ratio (which leads to the anomalous low O/H ratio using the strong-line abundance diagnostics), its metallicity is in fact comparable to that of the rest of the sample.

A least-squares fit to the data points (excluding the outlier \#14) using the 
PP04 calibration yields a radial galactocentric gradient \oh\,=\,$8.55 ~(\pm 0.03) - 0.014 ~(\pm 0.010)\, R/R_{25}$. The (virtually flat) gradient's slope is  unchanged using the D02 calibration, and can be expressed as d(O/H)/dR\,=\,$-0.009 ~(\pm 0.007)$ dex\,kpc$^{-1}$. The mean abundance values are $\langle$\oh$\rangle_{PP04}$ = $8.52 ~(\pm 0.03)$ and
$\langle$\oh$\rangle_{D02}$ = $8.63 ~(\pm 0.04)$, respectively. An additional indication of the 
gas metallicity
 is provided by the O3N2 = \{log{(\oiii\lin5007/\hbeta) $-$ N2\} diagnostic (PP04) based on \oiii\lin 5007 detections (triangles in Fig.~\ref{f3}, mean \oh\ = 8.65) and upper limits (arrows). Systematic offsets between different calibrations are well known to exist (\citealt{Kewley:2008}), and while their discussion is beyond the scope of this paper,  it is important to point out for the following discussion that the N2 diagnostic, together with the {\it direct} method, tends to yield lower O/H ratios compared to alternative strong-line abundance determination methods (\citealt{Bresolin:2009a}). In summary, the  diagnostics considered here indicate that the oxygen abundance in the outer disk of \ngc\ is essentially constant, and  I adopt a representative mean value \oh\,=\,8.6$\pm0.1$, equivalent to $\sim$0.8$\times$ the solar value (\citealt{Asplund:2009}). This result is consistent with the information provided by the photoionization models shown in Fig.~\ref{f2}.

\begin{figure}
\center \includegraphics[width=1.01\columnwidth]{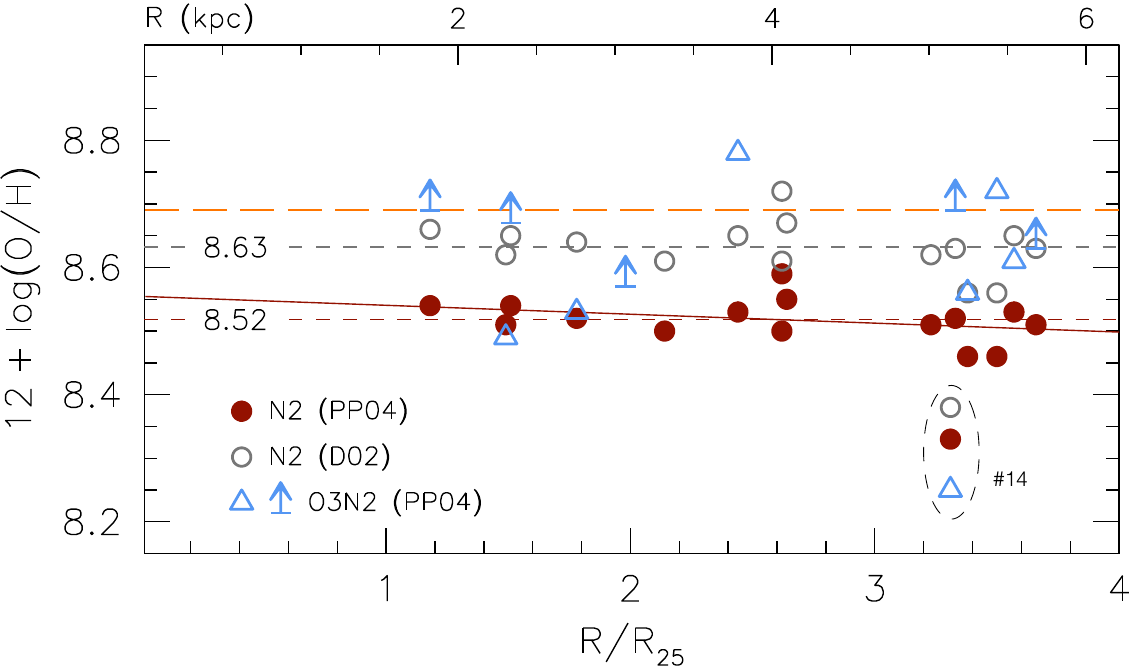}\medskip
\caption{Radial distribution of the O/H abundance ratio obtained from the N2 diagnostic, using the calibrations of 
\citet[=\,PP04]{Pettini:2004} and \citet[=\,D02]{Denicolo:2002}. The full line represents a least-squares fit to the data points measured with the PP04 calibration.
The  dashed lines show the mean values (indicated) of the O/H ratio resulting from the two calibrations.
Triangles and arrows correspond to the abundances measured from the O3N2 diagnostic based on \oiii\lin 5007
 detections and  upper limits, respectively. The solar O/H value is indicated by the upper dashed line.
\label{f3}}
\end{figure}

\bigskip
\section{Discussion}
The star formation history of \ngc\ has been recently investigated by \citet{Williams:2010} from theoretical fits to the color-magnitude diagrams (CMDs) of the stellar populations  in three  WFPC2 fields observed with the {\it Hubble Space Telescope}. These fields cover the galactocentric distance range 0.61 -- 2.55\,\rtf, overlapping with the  range spanned by the \hii\ regions observed in this work (1.06 -- 3.66\,\rtf).
The CMD analysis indicates that the majority (70\%) of the stars in the disk (even in the outer star forming ring)
 formed earlier than 10~Gyr ago (90\% formed by the end of the following 2~Gyr). The star formation rate (SFR) surface density declined from $\sim 6.8\times 10^{-3}$ \msun\,yr$^{-1}$ \,kpc$^{-2}$ early in the history of the galaxy to a negligibly small value 
($\lesssim 10^{-9}$ \msun\,yr$^{-1}$ \,kpc$^{-2}$)  
 approximately 0.9~Gyr ago, when the gas surface density was extremely small ($\sim 0.02$~\msun\,pc$^{-2}$). The CMD fitting suggests that in the  recent past ($\sim500$~Myr ago) the star formation activity
resumed, with an estimated SFR density of $2 \times 10^{-4}$ \msun\,yr$^{-1}$ \,kpc$^{-2}$ in the time interval between 0.5 and 0.2~Gyr ago (\citealt{Williams:2010}, Fig.~14). This  is about one order of magnitude higher than the current value  estimated from the 
far-UV emission of the star-forming ring (\citealt{Thilker:2010}). \citet{Williams:2010} speculated that the  increase in  star formation activity was possibly triggered by the merger 
postulated by \citet{del-Rio:2004}, who studied with 21\,cm observations the morphology and kinematics of the $1.5 \times 10^8$~\msun\ \hi\ outer disk.
\citet{del-Rio:2004} identified  the source of this gas  with a merging 
dwarf galaxy about 0.5-1~Gyr ago, lending credibility to the rejuvenation scenario to explain the recent star formation in \ngc.

Is this picture consistent with the chemical abundance analysis carried out in Sect.~3? It is worth comparing first  the present-day gas metallicity with the stellar metallicity 
in the disk, as inferred by \citet{Williams:2010} from the CMD modeling. As mentioned earlier, most of the galaxy's gas  converted into  stars very early on: the typical stellar age in \ngc\  is 12~Gyr, with a fitted metallicity [M/H]\,=\,$-0.75\pm0.37$. The declining  star formation activity caused a progressive metal enrichment of the galaxy. The errors in [M/H] given by \citet{Williams:2010} become too large ($>1$~dex) at recent times ($<200$~Myr ago) to offer a meaningful comparison with the \hii\ region chemical abundances, but these authors estimated a stellar metallicity [M/H]\,=\, $-0.38 \pm 0.35$ about 750~Myr ago. This is consistent with the value  
[M/H]\,$\simeq$\,$-0.4$ measured by \citet{Seth:2010} from  spectra of the \ngc\ bulge (representative age: 5~Gyr). These results suggest that 
a chemical abundance comparable to the one obtained for the outer ring \hii\ regions, 
[O/H]\,$\simeq -0.1 \pm 0.1$, was likely reached in the disk of this galaxy at least 1 Gyr ago (before the postulated merger),  considering that for typical S0 galaxies [$\alpha$/Fe]\,$\simeq$\,0.3 (\citealt{Silchenko:2012}; the \hii\ regions provide a determination of the O/H ratio, while the stellar M/H given above is assumed to be representative of the Fe/H ratio).

If the general description of the star formation history of \ngc\ outlined by \citet{Williams:2010}
is broadly correct, the significant chemical enrichment measured for the outer disk \hii\ regions
argues against a gas-rich dwarf galaxy as the main source of fuel responsible for the re-ignition of the star formation, as proposed by \citet{del-Rio:2004} and \citet{Thilker:2010}. 
From the stellar mass of \ngc\ measured by \citet{Thilker:2010},
and assuming a mass merger ratio of 1:3 (the maximum ratio commonly adopted to distinguish between major and minor mergers), the stellar mass of the merging dwarf would be $1.5\times 10^8$\,\msun. According to the mass-metallicity relation observed in the local universe (\citealt{Berg:2012}) a galaxy with this mass has an oxygen abundance \oh\,$\simeq$\,$7.98\pm0.15$.\footnote{The gas metallicities obtained from the N2 method used here are on the same absolute scale as 
the {\it direct} metallicities measured by \citet{Berg:2012}.}

Allowing for star formation over a period of 0.3 Gyr at a rate of  $2 \times 10^{-4}$ \msun\,yr$^{-1}$ \,kpc$^{-2}$ (following \citealt{Williams:2010}), considering an observed \hi\ surface density in the star-forming ring $\Sigma_{H\,I} = 1$~\msun\,pc$^{-2}$ (\citealt{del-Rio:2004}) and with an initial 
oxygen abundance \oh\,$\simeq$\,7.98, I estimate a final abundance of only \oh\,$\simeq$\,8.16, considering an oxygen yield of 0.01 (see Eq.~1 in \citealt{Bresolin:2012}). Even if we double this period of enhanced star formation to 0.6~Gyr (which is excluded by the star formation history of the galaxy presented by \citealt{Williams:2010}), the present-day oxygen abundance would be \oh\,$\simeq$\,8.30, still 0.3 dex below the observed value. Even considering the uncertainty in the estimated mass of the merging dwarf, the scatter in the mass-metallicity relation and the uncertainty in the nebular O/H resulting from  the use of the N2 diagnostic (\citealt{Perez-Montero:2009a}), it appears difficult to explain the  chemical enrichment of the outer disk of \ngc\ with the proposed dwarf galaxy merger. 

The relatively high nebular metallicity of the star-forming ring of \ngc\ also rules out (metal-poor) primordial IGM gas accretion as the primary source of its star-forming gas (which has been proposed as a viable mechanism for the formation of metal-poor  galactic ring structures, e.g.~by \citealt{Finkelman:2011} and \citealt{Spavone:2010}). On the other hand, 
a near-solar metallicity would be expected if  the bulk of the gas  has been part of the galaxy since its early history, and  has been progressively chemically enriched by the star formation that has occurred in the disk (in agreement with the declining star formation scenario discussed by \citealt{Salim:2012} and \citealt{Fang:2012}), although this possibility was deemed unlikely by \citet{del-Rio:2004}, based on 
the  warped nature of the \hi\ structure, which suggests that the gas has been acquired  recently,
and on critical gas density arguments (invalidated by the GALEX detection of recent star formation). It is  conceivable that a perturbing event (minor/dry merger) took place in the recent past, that revitalized the star formation activity, but it is unlikely, based on the reconstructed star formation history, that such an event was responsible for the accretion of the gas, because of the incompatible chemical composition.
It is also possible to argue in favor of an enriched accretion mechanism, as  proposed in  recent studies of the chemical abundances of the outer disks of spiral galaxies (\citealt{Bresolin:2009, Bresolin:2012}; but I also note that these structures are 3$\times$ larger than the outer disk of \ngc), where the gas, chemically enriched by stellar processes in the disk, 
is subject to a wind-recycling mechanism, and inflows back into the disk (\citealt{Dave:2011, Martin:2012}).

In conclusion, this work shows that the chemical abundance analysis of the ionized gas provides crucial constraints 
on the origin of the outer star forming rings and potential rejuvenation events observed in ETGs. In particular, two of the main scenarios proposed,
cold accretion and  minor merger, are ruled out in the case of \ngc\ by the  high nebular metallicity observed.
Future work will test whether the declining star formation scenario provides a better fit to the  present-day gas metallicity of additional ETGs.

\acknowledgments
FB gratefully acknowledges the support from the National Science Foundation grant AST-1008798, and thanks
the anonymous referee for constructive comments.

\clearpage
% - - - - - - - - - - - - - - - - - - - - - - - - - - - -
%\bibliography{/Users/fabio/PDF-Papers/Papers}

\end{document}